\documentclass[a4paper,
               biblatex,     
               ]{jacow}
\usepackage{pdfpages,multirow,ragged2e} %
\makeatletter%
	\ifboolexpr{bool{xetex}}
	 {\renewcommand{\Gin@extensions}{.pdf,%
	                    .png,.jpg,.bmp,.pict,.tif,.psd,.mac,.sga,.tga,.gif,%
	                    .eps,.ps,%
	                    }}{}
\makeatother

\ifboolexpr{bool{xetex} or bool{luatex}} 
 {}                                      
 {\usepackage[utf8]{inputenc}}           

\usepackage[USenglish]{babel}
\usepackage{subfig}
\usepackage{graphicx}

\DeclareSIUnit\bar{bar} 

\ifboolexpr{bool{jacowbiblatex}}%
 {
  \addbibresource{MOPA132.bib}
 }{}
\title{Status Report on the
\\Hydrodynamic Simulations of a Tapered Plasma Lens
\\for Optical Matching at the ILC \NoCaseChange{e\textsuperscript{+}} Source
\thanks{Talk presented at the International Workshop on Future Linear Colliders (LCWS 2023), 15-19 May 2023. C23-05-15.3.}}

\author{M.~Formela\thanks{manuel.formela@desy.de},  N.~Hamann, G.~Moortgat-Pick\textsuperscript{1}, II. Institute of Theoretical Physics, \\ University of Hamburg, Hamburg, Germany \\
G.~Loisch, M.~Mewes, M.~Thévenet, J.~Osterhoff, Deutsches Elektronen-Synchrotron DESY, \\ Hamburg, Germany \\
G.~Boyle, James Cook University, Townsville, Australia \\
\textsuperscript{1}also at Deutsches Elektronen-Synchrotron DESY, Hamburg, Germany}
\begin{document}
\maketitle
\begin{abstract}
The International Linear Collider is a planned electron-positron linear collider with its positron source producing positrons by aiming undulator radiation onto a rotating target.
The resulting, highly divergent positron beam requires immediate optical matching to improve the luminosity and therefore the success of the intended collision experiments.
Here, optical matching refers to the process of capturing particles and making them available for downstream beamline elements like accelerators.
In the past, this has been done with sophisticated coils, but more recently the usage of a current-carrying plasma, a so-called plasma lens, has been proposed as an alternative.
For the International Linear Collider idealised particle tracking simulations have already been done with the purpose of finding the optimal plasma lens design with respect to the captured positron yield.
The proposed design is characterised by a linearly widened radius in beam direction~\cite{Formela:2022gco}.
Now further research and development of this design is required, including both experiments with a prototype set-up as well as corresponding simulations modelling the hydrodynamics of the current-carrying plasma and the resulting magnetic field.
The accuracy of the latter will benefit greatly from the former.
In this work, first preliminary hydrodynamic simulations instil confidence into further endeavours. 
\end{abstract}
\section{Introduction}
The plasma lens serves as a device for focusing charged particles by utilising a current-carrying plasma and its potential applications include final focusing and particle matching.
When this focusing is externally driven, typically by a laser or a power supply, it is referred to as an active plasma lens (APL).
Typically, an APL consists of a capillary aligned along the beam axis, accompanied by lateral gas inlets.
For APLs driven by a power supply, electrodes are added to both ends of the capillary.\\
The operation of such an APL begins with filling the empty capillary with gas (e.g., H\textsubscript{2} or Ar) via the inlets.
Subsequently, a high voltage is applied between the capillary's electrodes, generating an electric field within the capillary.
If the electric field is sufficiently strong, it ionises some of the molecules near the electrode edges, producing free electrons.
These free electrons are then accelerated by the electric field, gaining enough energy to create additional free electrons through impact ionisation.
This process continues, leading to a chain reaction resulting in a partially or fully ionized plasma.
During this process, the primarily longitudinal electric current significantly increases and induces an azimuthal magnetic field.
This azimuthal component causes the incoming charged particle beam to undergo axial symmetric deflection, either focusing or defocusing the beam depending on the relative orientation between the field, particle motion, and charge polarity.
While this provides a basic understanding of APL functionality, more advanced processes, such as plasma pinching, have been omitted here.\\
In theory, APLs should have advantages over conventional optical matching devices, primarily due to their axial symmetric focusing properties created by the azimuthal magnetic field.
However, three major challenges must be addressed for the International Linear Collider (ILC).\\
First, the plasma lens must consistently match successive positron pulses.
In the case of the ILC, the positron beam's temporal structure (see Table~\ref{tab:e+_beam_structure}) imposes demanding requirements on the discharge pulse's timing and duration, balancing low plasma disturbance with consistent focusing and technical feasibility by the power supply.\\
The second challenge is the potential gas outflow from the windowless APL capillary into the downstream accelerator, risking vacuum contamination and discharges within the accelerator.\\
And lastly, the ability of APL components to withstand the collective heat load from electric heating and particle beam deposition must be examined, as excessive heat could deform the plasma lens setup or lead to gas contamination.
\begin{table}[h!]
\centering
    \caption{Positron Beam Structure with $\num{1312}$ Bunches per Pulse.}
   \begin{tabular}{lccc}
       \toprule
       & \textbf{Repetition rate} & \textbf{Duration} & \textbf{Spacing} \\
       \midrule
    Bunch train & $\qty{5}{\Hz}$ & $\qty{727}{\mu\s}$ & $\qty{199}{\ms}$ \\
    Bunch & $\qty{1.8}{\MHz}$ & $\qty{538}{\ps}$ & $\qty{554}{\ns}$ \\ 
       \bottomrule
   \end{tabular}
   \label{tab:e+_beam_structure}
\end{table}
\subsection{Particle Tracking}
\begin{figure}[h!]
    \centering
    \includegraphics[scale=0.25]{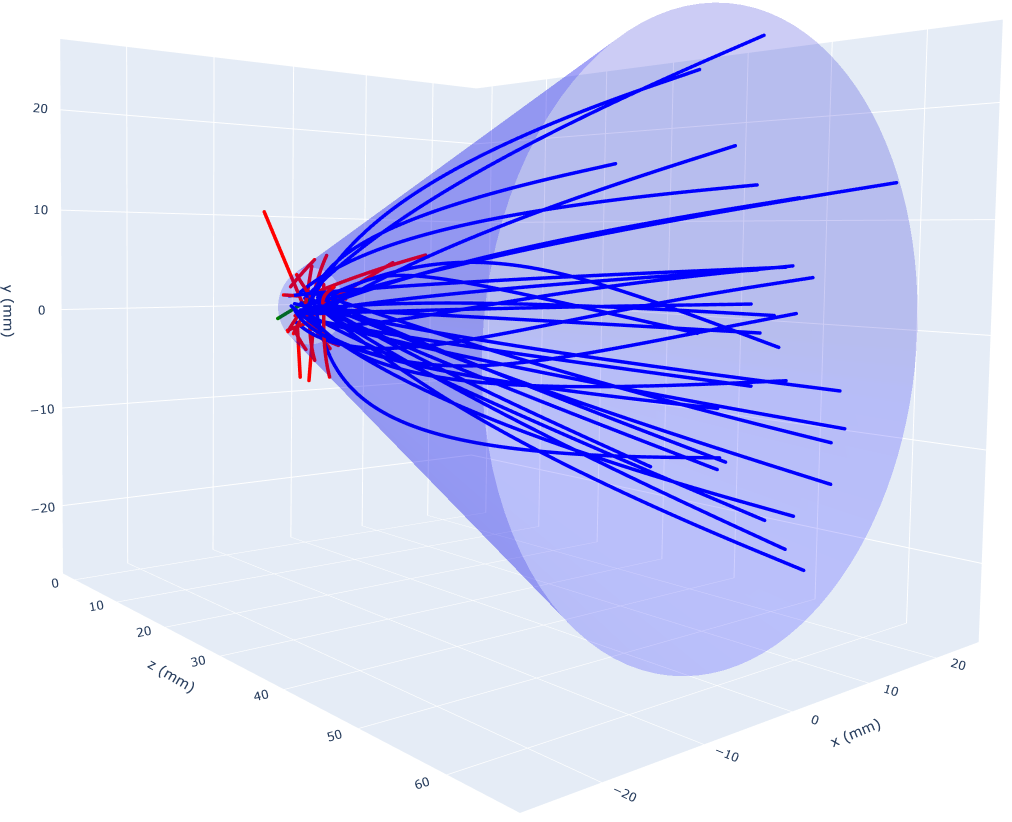}
    \caption{Visualisation of particle tracking in a cone-shaped plasma lens.}
    \label{fig:particle_tracking-example}
\end{figure}

In previous work towards an APL for the ILC e+ source, APL design parameters were optimised to maximise the number of matched positrons.
Particle tracking simulations were conducted without simulating plasma dynamics, assuming an idealised plasma lens with a purely longitudinal and radially homogeneous electric current density
\begin{equation*}
    \vec j(x,y,z) = \vec j_z(z).
\end{equation*}
According to the Maxwell equations, the induced magnetic field is therefore
\begin{equation*}
    B(\rho,\varphi,z) = B_\varphi(\rho,z) = \begin{aligned}
    &\frac{\mu_0 I}{2\pi R(z)^2}\rho,  && \text{for} \ 0 \leq \rho \leq R(z)
  \end{aligned}
\end{equation*}
with the vacuum permeability~$\mu_0$, electric current~$I$ and capillary radius~$R$.\\
The optimisation process was conducted with the particle tracking code ASTRA~\cite{floettmann2017astra} and resulted in an APL design characterised by its conical shape (see Fig.~\ref{fig:particle_tracking-example}), capable of a positron capture efficiency of approximately $\qty{43}{\percent}$~\cite{Formela:2022gco}, which is significantly higher than the currently proposed ILC optical matching device, the quarter-wave transformer~\cite{Gai:418711}.
The design also past stability tests successfully, where parameter errors of $\pm\qty{10}{\percent}$ were introduced and which resulted in a relative decline in the number of captured particles by at most $\qty{5}{\percent}$~\cite{Formela:2022gco}.
However, this design is preliminary and will be further optimised as more accurate magnetic field information becomes available through hydrodynamic simulations.
\subsection{Downscaled Prototype}
Building and measuring the proposed APL design in experiments is challenging due to demanding power supply requirements and the large size of the APL.
To enable experiments, the design geometry was downscaled by a factor of approximately five while keeping the electric current density constant.
The downscaled prototype design parameters are provided in Table~\ref{tab:optimal_downscaled_design}.
More details on the downscaled prototype and planned future experiments can be found in Ref.~\cite{Hamann:2023gco}.
\begin{table}[h!]
\centering
    \caption{Parameters of the Prototype Plasma Lens Design.}
   \begin{tabular}{lccc}
       \toprule
       \textbf{Parameter name} & \textbf{Symbol} & \textbf{Value} & \textbf{Unit} \\
       \midrule
    Electric Current & $I_0$ & $\num{350}$ & $\unit{\A}$ \\
    Tapering Type & & linear & \\
    Opening Radius & $R_\mathrm{0}$ & $\num{0.85}$ & $\unit{\mm}$ \\
    Exit Radius & $R_\mathrm{1}$ & $\num{5.0}$ & $\unit{\mm}$ \\
    Tapering Length & $L$ & $\num{12}$ & $\unit{\mm}$ \\
       \bottomrule
   \end{tabular}
   \label{tab:optimal_downscaled_design}
\end{table}
\section{HYDRODYNAMIC SIMULATIONS}
In the future, comprehensive hydrodynamic (HD) simulations of both the downscaled and full-scale APL designs are planned using COMSOL Multiphysics\textsuperscript{\tiny\textregistered}~\cite{COMSOL_Multiphysics}.
Currently, only the downscaled prototype has been simulated, focusing on hydrodynamics while neglecting magnetic effects.
The HD simulations are based on a custom model treating the hydrogen plasma as a quasi-neutral, two-temperature, reacting homogeneous fluid.
This model accounts for separate temperatures for electrons and heavier particles (ions, neutrals) and calculates plasma composition using diffusion and reaction rates.\\
The prototype geometry was modelled in 2D, with axial symmetry assumed for the 3D simulation.
The model simplified the geometry, assuming electrodes spanning the capillary openings, perfectly insulating capillary walls and a sine-like half-wave discharge current with an amplitude of $\qty{350}{\A}$ and a duration of $\qty{200}{\ns}$.
\subsection{Results}
\begin{figure}
  \centering
  \subfloat[a][At APL entrance]{\includegraphics[scale=0.38]{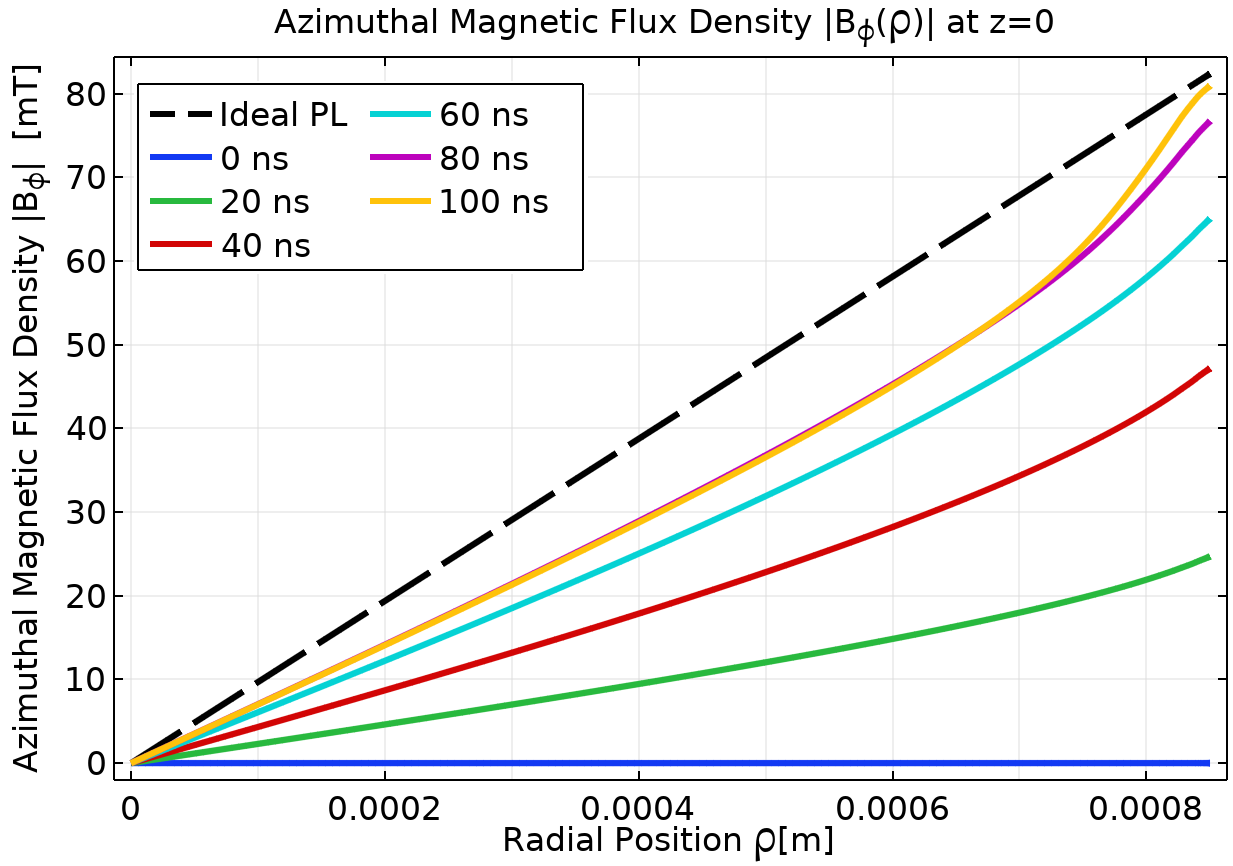} \label{fig:Bphi-z=0}} \\
  \subfloat[b][At APL exit]{\includegraphics[scale=0.38]{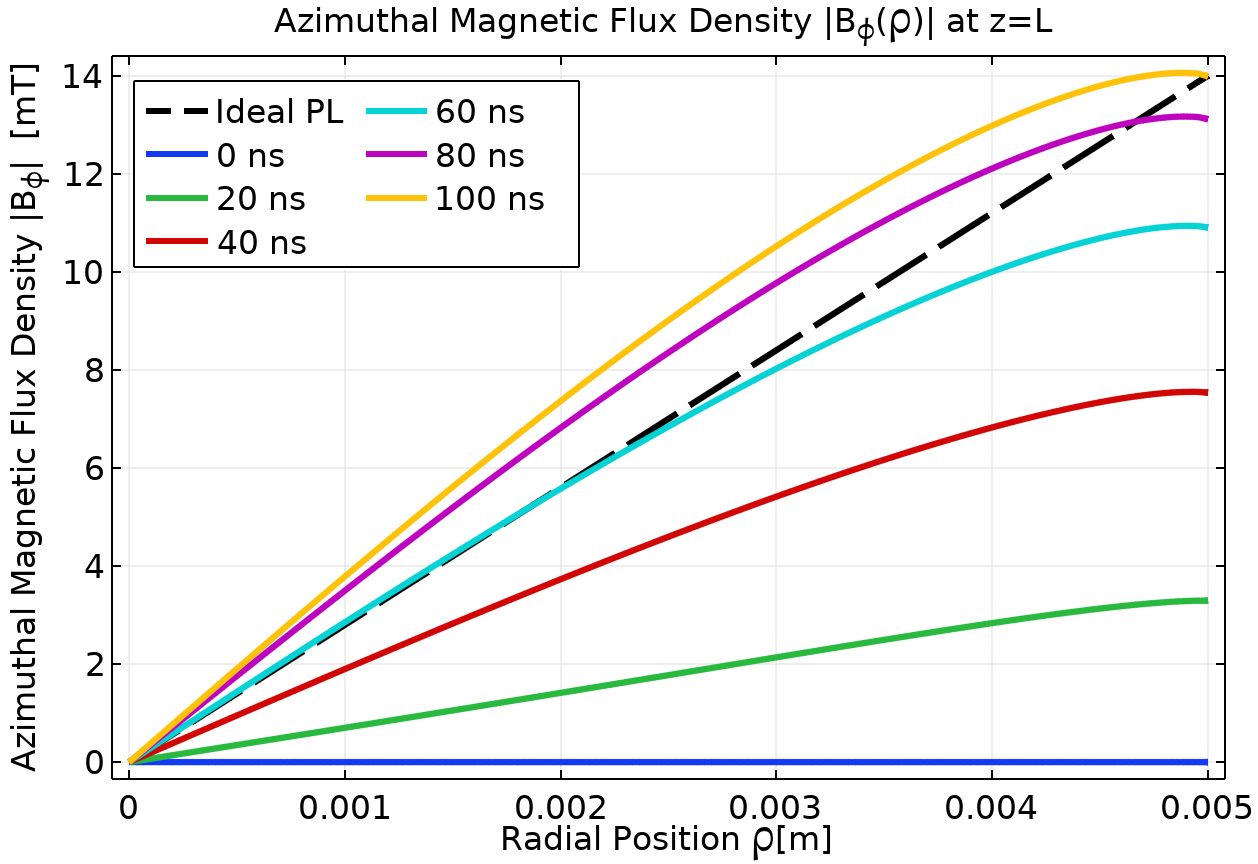} \label{fig:Bphi-z=L}}
  \caption{Radial distribution of the azimuthal magnetic flux density. Solid: HD simulation at various times; Dashed: Ideal model with homogeneous current density.}
  \label{fig:Bphi}
\end{figure}

Fig.~\ref{fig:Bphi} shows the azimuthal magnetic flux density~$|B_\varphi(\rho)|$ plotted against the radial position~$\rho$ at the entrance and exit of the APL, respectively.
The coloured, solid lines correspond to the HD simulation results at different times, while the dashed, black line belongs to the ideal plasma lens. \\
According to Fig.~\ref{fig:Bphi-z=0}, the magnetic field at the entrance is at most approximately $\qty{20}{\percent}$ lower for the HD simulation compared to the ideal APL.
While this deviation is not expected to be problematic due to the design's inherent margin for matched positrons, further investigations through simulations and experiments are required.
In contrast, Fig.~\ref{fig:Bphi-z=L} shows that the situation is reversed at the APL exit.
\section{OUTLOOK}
Future investigations will focus on the plasma's response to multiple successive plasma discharge pulses, optimising the discharge pulse structure to meet the demanding requirements of the positron beam's temporal structure.%
Gas flow measurements from the APL capillary into the downstream accelerator will be conducted, and the impact of the collective heat load on APL components will be examined.

To improve the simulation model's accuracy, various enhancements are planned, including the use of argon instead of hydrogen, more realistic electrodes and discharge pulses, and the addition of inlets.
The simulated magnetic field data will be incorporated into particle tracking simulations to further optimise the plasma lens shape for maximising the number of captured positrons.
Additionally, experiments on the downscaled prototype will be conducted at the ADVANCE Laboratory at DESY to validate simulation results and ensure the success of the plasma lens development.
\section{ACKNOWLEDGEMENTS}
This project received funding from the German Federal Ministry of Education and Research [Grant No.~05P21GURB1]. \\
Also, we express our deeply felt gratitude towards K. Fl\"ottmann and M. Fukuda for multiple instances of valuable collaboration.
%
%
%

\ifboolexpr{bool{jacowbiblatex}}%
	{\printbibliography}%

\end{document}